\documentclass{aa}  
\usepackage{graphicx}
\usepackage{txfonts}
\newcommand{\mincir}{\raise -2.truept\hbox{\rlap{\hbox{$\sim$}}\raise5.truept
\hbox{$<$}\ }}
\newcommand{\magcir}{\raise -2.truept\hbox{\rlap{\hbox{$\sim$}}\raise5.truept
\hbox{$>$}\ }}
\newcommand{\siml}{\raise -2.truept\hbox{\rlap{\hbox{$\sim$}}\raise5.truept
\hbox{$<$}\ }}
\newcommand{\simg}{\raise -2.truept\hbox{\rlap{\hbox{$\sim$}}\raise5.truept
\hbox{$>$}\ }}
\newcommand{\be}{\begin{equation}}
\newcommand{\ee}{\end{equation}}
\newcommand{\ba}{\begin{eqnarray}}
\newcommand{\ea}{\end{eqnarray}}

\newcommand {\h} {$h_{70}^{-1}$ Mpc$\;$}

\newcommand {\hhh} {\;h_{70}^{-1} \mathrm{Mpc}}

\newcommand {\kss} {km~s$^{-1}$}

\begin{document}
   \title{Environmental effects on the bright end of the 
galaxy luminosity function in galaxy clusters}
   \author{
R. Barrena\inst{1,2}
          \and
M. Girardi\inst{3,4}
          \and
W. Boschin\inst{5}
          \and
F. Mardirossian\inst{3,4}
}

   \offprints{R. Barrena, \email{rbarrena@iac.es}}

   \institute{ 
     Instituto de
     Astrof\'{\i}sica de Canarias, C/V\'{\i}a L\'actea s/n, E-38205 La
     Laguna (Tenerife), Canary Islands, Spain\\ 
\and Departamento de
     Astrof\'{\i}sica, Universidad de La Laguna, Av. del
     Astrof\'{\i}sico Francisco S\'anchez s/n, E-38205 La Laguna
     (Tenerife), Canary Islands, Spain
\and Dipartimento di Fisica dell' Universit\`a degli Studi
     di Trieste - Sezione di Astronomia, via Tiepolo 11, I-34143
     Trieste, Italy\\ 
\and INAF - Osservatorio Astronomico di Trieste,
     via Tiepolo 11, I-34143 Trieste, Italy\\ 
\and Fundaci\'on Galileo
     Galilei - INAF, Rambla Jos\'e Ana Fern\'andez Perez 7, E-38712
     Bre\~na Baja (La Palma), Canary Islands, Spain\\ 
}

\date{Received  / Accepted }

\abstract{The dependence of the luminosity function (LF) of cluster
  galaxies on the evolutionary state of the parent cluster is still an
  open issue, in particular as concern the formation/evolution of the
  brightest cluster galaxies.} {We plan to study the bright part of
  the LFs of a sample of very unrelaxed clusters (``DARC'' clusters
  showing evidence of major, recent mergers) and compare them to a reference
  sample of relaxed clusters spanning a comparable mass and redshift
  range.} {Our analysis is based on the SDSS DR7 photometric data of
  ten, massive, and X-ray luminous clusters ($0.2\lesssim z \lesssim
  0.3$), always considering physical radii ($R_{200}$ or its
  fractions). We consider $r'$ band LFs and use the
  color-magnitude diagrams ($r'-i'$,$r'$) to clean our samples as well
  to consider separately red and blue galaxies.}{We find that DARC and
  relaxed clusters give similar LF parameters and blue fractions.  The
  two samples differ for their content of bright galaxies BGs,
  $M_{r'}<-22.5$, since relaxed clusters have fewer BGs, in particular
  when considering the outer cluster region $0.5R_{200}<R<R_{200}$ (by a
  factor two). However, the cumulative light in BGs is similar for
  relaxed and DARC samples.}{We conclude that BGs grow in luminosity
  and decrease in number as the parent clusters grow hierarchically in
  agreement with the BG formation by merging with other luminous
  galaxies.}


   \maketitle
%

\section{Introduction}
\label{intr}

The study of the formation and evolution of galaxies is a fundamental
avenue of research in the process of understanding astrophysical and
cosmological issues.  The galaxy luminosity function (LF) -- the
number of galaxies per unit volume in the luminosity interval $L$ and
$L+dL$ -- is one of the most direct observational test of theories of
galaxy formation and evolution.

Clusters of galaxies are ideal systems within which to measure the
galaxy LF for the large number of galaxies at the same distance. On
the other hand, clusters represent an extreme environment for galaxy
evolution, either in situ or through the accretion of galaxies within
groups, which are situated in the filamentary structure of the
hierarchical Universe (e.g., Poggianti et al. \cite{pog05}).

While introducing the modern form of the LF, the so--called
``Schechter Function'', Schechter (\cite{sch76}) suggested that the
cluster LF is universal in shape (with a turnover at
$M_B^*=-20.6+5{\rm log} h_{50}$ and a faint-end slope of
$\alpha=-1.25$). Much progress has then been made in looking for
possible observational signatures of a non universality of the cluster
LF. That different morphological types are characterized by different
LFs (Binggeli et al. \cite{bin88}) and the well known morphological
segregation with the local density and thus the clustercentric radius
(Dressler \cite{dre80}; Whitmore \cite{whi91}) make more complex the
study of the cluster LF.  In this context, several studies about LF
concern the possible variation of the cluster LF with the local
projected galaxy density and/or cluster-centric radius. In particular,
there are claims for a radius-dependent steepening of the galaxy LF at
faint magnitudes, likely due to red galaxies (e.g., de Propris
\cite{dep95}; Driver et al. \cite{dri98}; Popesso et al. \cite{pop06};
Barkhouse et al. \cite{bar07}). However, the few cases of LFs based on
deep spectroscopy exhibit shallower slopes (e.g., Rines \& Geller
\cite{rin08} and refs.  therein) suggesting that the field
contamination might be the cause of the observed steep LFs based on
photometry only.  Thus, the question is still far from a definitive
conclusion.
  
Another issue concerns the possible dependence of the LF on global
cluster properties.  Some studies interest a few well studied clusters
and/or cluster complexes (e.g., Abell 209 in Mercurio et al. \cite{mer03},
Coma vs. Abell 1367 in Iglesias-P\'aramo et al. \cite{igl03}) and other studies
use instead a large statistical base. In particular, there are
indications in favour of correlations between the LF parameters and the
cluster mass or its proxies (e.g., richness, luminosity, galaxy
velocity dispersion).  For instance, Valotto et al. (\cite{val97})
find that poorer clusters have a flatter LF.  But, when the LF is
properly calculated within the cluster physical sizes, given by
$R_{200}$ or $R_{500}$, the correlation between the dwarf-to-giant
ratio and X-ray (or optical) luminosity disappear (cf. Popesso et
al. \cite{pop06} with Popesso et al. \cite{pop04}). This suggests that
the radial behavior of the LF should be taken into account to obtain
unbiased conclusions. However, also when considering an appropriate
cluster physical size, the conclusions are not all consistent. For example,
Hilton et al. (\cite{hil05}) find that clusters with lower X--ray
luminosity have a brighter $M^*$, but Barkhouse et al. (\cite{bar07})
find no correlation.  Recent studies also attempt to analyze more
distant clusters, but again with no conclusive results (Gilbank et
al. \cite{gil08}; Rudninick et al. \cite{rud09} and discussion
therein).

Possible correlations with the presence/absence of a cD galaxy or with
the Bautz-Morgan type (the BM-type classification based on the
brightest cluster galaxies, BCGs, Bautz \& Morgan \cite{bau70}), are,
maybe, also more interesting for their direct connection with the
cluster evolutionary state. In fact, BCGs or cD galaxies are expected
to be created i) via the merger of giant galaxies in an early phase
during the cluster collapse (Merritt \cite{mer85}) or ii) in a
following phase due to the dynamical friction acting on late-comer
galaxies (e.g., Ostriker \& Tremaine \cite{ost75}) or iii) via the
disruption and cannibalization of many faint galaxies in the cluster
cores (L\'opez-Cruz et al. \cite{lop97}). The first two mechanisms
might reduce the number of bright galaxies, thus shifting $M^*$ to a
fainter value and, indeed, Barkhouse et al. (\cite{bar07}) find a
weak correlation with the BM-type (but see de Propris et
al. \cite{dep03} for no correlation). Another consequence can be the
increment of the gap between the first and the second brightest
galaxies in each cluster and, in fact, this magnitude gap is shown to
be larger in clusters without substructure and smaller in clusters
with substructure (Ramella et al. \cite{ram07}).

As for the second mechanism, L\'opez-Cruz et al. (\cite{lop97}) find
that seven rich, massive, cD clusters with symmetric X-ray morphology
show a flat faint-end slope while a steeper faint-end slope is
detected in poorer clusters but also in a rich, binary cluster like
the Coma cluster.  Somewhat in agreement, Driver et al. (\cite {dri98})
find that the dwarf-to-giant ratio increases with larger BM-types (i.e.,
likely less evolved clusters).  In this context, Valotto et
al. (\cite{val04}) pointed out the importance of the projection
effects in determining the faint-end slope. In fact, in the large cluster
sample analyzed by Valotto et al. it is the X-ray selection and not 
the cluster domination by central galaxies what determines the flatness 
of the faint-end slope thus suggesting that real, compact clusters are
less affected by the field contamination.  However, although this bias
could explain possible differences between rich and poor clusters, the reason
for the difference between the seven cD relaxed clusters of L\'opez-Cruz
et al. (\cite{lop97}) and the Coma cluster is not clear yet.

As for the possible dependence of the LF on the dynamical status of
the cluster, the noticeable study of de Propris et al. (\cite{dep03})
analyze 60 clusters and find that the LF is similar for clusters with
and without substructure. Very few detailed works are devoted to study
the LFs of subclumps in individual clusters to find possible evidence
of cluster mergers (e.g., Durret et al. \cite{dur11}).

As reported above, it is particularly interesting the bright end of the
LF since there is a long debate on the BCG formation.  For instance,
semianalytical models suggest that BCGs assemble surprisingly late
with the stars formed very early in many small galaxies (e.g., 50\% at 
z=0.5, 80\% at z=0.3; see de Lucia \& Blaizot \cite{del07}).
However, photometric and chemical observables suggest that it is
difficult to explain giant elliptical by a pure sequence of multiple
minor dry mergers or via major dry mergers (e.g., Stott et
al. \cite{sto10}; Ascaso et al. \cite{asc11}) and numerical
simulations are continuously improved to reproduce them better
(e.g., Ruszkowski \& Springel \cite{rus09}). Moreover, galaxy-galaxy
mergers are most efficient within small halos with small velocity
dispersion and the merger of BCGs in clusters requires a long time 
(Dubinski \cite{dub98}). However, the correlation between BCG luminosity
and parent cluster mass strongly suggest that BCGs co-evolves with 
the cluster (Lin \& Mohr \cite{lin04}).

In this context, we plan to study the LFs of a sample of massive
and very unrelaxed clusters and compare them to a reference sample
of relaxed clusters spanning a comparable mass and redshift
range.

As for the unrelaxed clusters, we are going on with a long term
project to study clusters exhibiting large, diffuse radio sources,
i.e. radio halos and relics, based on spectroscopic data mostly
acquired at TNG (DARC - Dynamical Analysis of Radio
Clusters\footnote{other information are given in the web site
  http://adlibitum.oat.ts.astro.it/girardi/darc.}  - e.g., Girardi et
al. \cite{gir07}).  These radio sources are rare phenomena and are due
to the synchrotron radio emission of widespread relativistic particles
embedded in the cluster magnetic field (e.g., see Feretti et
al. \cite{fer02b} for a review).  Cluster mergers have been proposed
to provide the large amount of energy necessary for electron
reacceleration to relativistic energies and for magnetic field
amplification (e.g., Tribble \cite{tri93}; Ensslin et
al. \cite{ens98}; Brunetti et al. \cite{bru09}) or required to explain
the time-dependence of the magnetic fields in recent models based on
secondary electrons (Keshet \& Loeb \cite{kes10}; Keshet
\cite{kes11}).  Indeed, from the observational point of view, there is
growing evidence of the connection between diffuse radio emission and
cluster merging, since up to now diffuse radio sources have been
detected only in merging systems.  In most past studies the cluster
dynamical state was derived from X--ray observations (see Buote
\cite{buo02}; Feretti \cite{fer06} and \cite{fer08}; Cassano et
al. \cite{cas10}).  In agreement, our analyses, based on optical data
and the kinematics of member galaxies, detect strong subclustering and
find evidence of cluster mergers.  Summarizing, clusters with radio
halos/relics are the ideal cases to study the effect of the
cluster formation on the LF.

Here we analyze a subsample of five DARC clusters we have analyzed in the
past and all having available photometry in the the Sloan Digital Sky
Survey (SDSS). The DARC clusters form our sample of unrelaxed
clusters.  For comparison, we analyze a sample of relaxed clusters of
comparable mass and redshift.  Considering the magnitude limit of SDSS
photometry, the typical redshift of the samples, $z\sim 0.25$, limits
our study to the bright part of the LF. We plan to extend this study
to a fainter magnitudes in the next future.

Unless otherwise stated, we give errors at the 68\% confidence level
(hereafter c.l.).

Throughout this paper, we use $H_0=70$ km s$^{-1}$ Mpc$^{-1}$ in a
flat cosmology with $\Omega_0=0.3$ and $\Omega_{\Lambda}=0.7$.  We
define $h_{70}=H_0/(70$ km s$^{-1}$ Mpc$^{-1}$).

\section{Cluster sample}
\label{sample}

Among the DARC clusters of which we have already analyzed the internal
dynamics using a large number of member galaxies, five clusters (with
62-113 spectroscopic member galaxies) have also available photometry in 
the SDSS (Data Release 7): Abell~697, Abell~773, Abell~959, Abell~1240, and
Abell~2219, hereafter A697, A773, A959, A1240, A2219 (see Table~1 for
the respective reference sources). Their redshifts span the range
$z=0.19-0.29$ and they are characterized by high line-of-sight galaxy
(LOS) velocity dispersion $\sigma_{\rm v}$, X--ray temperature $T_{\rm
  X}$, and mass $M$ (see Table~\ref{tabdarc}). Evidence for their
unrelaxed dynamical state is reported in our previous studies.  In
particular, they span different angle of views for the merging
axis. For instance, the cluster merger occurred largely in the plane
of the sky in A1240 and largely along the LOS in A773.

As for comparison, we select from the literature a few well known
examples of very relaxed clusters spanning a similar $z$ range
($0.2 \lesssim z \lesssim 0.3$), 
characterized by high $T_{\rm X}$, and sampled by SDSS.
From the relaxed clusters listed by Allen et
al. (\cite{all04}) we select\footnote{Note that we exclude Abell~611
  due to its unrelaxed appearance in the SDSS image.} Abell~383,
Abell~963, and Abell~1835 (hereafter A383, A963, A1835). These
clusters are all classified as cool core clusters according to Baldi
et al. (\cite{bal07}). This agrees with that observations suggest that
cool cores are destroyed by cluster mergers and cool core clusters are
in the final phase of cluster relaxation (see e.g., Allen et
al. \cite{all01}; Buote \cite{buo02}; Sanderson et al. \cite{san06}).
Among the cool core clusters listed by Baldi et al. (\cite{bal07}) we
considered other two cool core clusters to complete our comparison
sample: Abell 2261 (hereafter A2261) and ZwCl 1021.0+0426 (hereafter
and better known as ZwCl3146).

To avoid the inherent bias that has plagued numerous studies (see the
section above), the cluster LFs will be compared based on scaling
relative to the dynamical radius, $R_{200}$. As an estimate of
$R_{200}$, we use $R_{\rm vir}$ definition by Girardi \& Mezzetti
(\cite{gir01}), Eq.~1 with the scaling of $H(z)$, 

\begin{equation}
R_{\rm 200}=0.17\times \sigma_{\rm v}/({\rm  km\ s}^{-1})/H(z)\hhh,
\end{equation}

\noindent where $\sigma_{\rm v}$ is the LOS velocity
dispersion.  The same result was obtained by Carlberg et
al. (\cite{car97}, see their Eq.~ 8 for $R_{200}$) using a different,
more theoretical, approach.  Biviano et al. (\cite{biv06}) obtained a
$10\%$ smaller estimate of $R_{200}$, $R_{\rm v}=0.15\times
\sigma_{\rm v}/({\rm km\ s}^{-1})/H(z)$ \h on the base of N-body
simulations. We also consider internal and external regions, i.e.  $R<
0.5 R_{\rm 200}$ and $0.5 R_{\rm 200}<R< R_{\rm 200}$,
respectively.  The centers for relaxed clusters are taken from X-ray
studies, while those for DARC clusters are those used in our previous 
papers. Note that, due to the unrelaxed state of DARC clusters, the
choice of the cluster center is not obvious since gas and galaxy
spatial distributions are generally different. For A697, A773 
and A2219 clusters we used the position of the brightest dominant galaxy,
very close to the X-ray centroid (or X-ray main peak), while for A959 
and A1240 we use X--ray centroid. However, in these two cases, X--ray 
centroid is located in between the two brightest dominant galaxies. A 
radius of $0.5R_{200}$ is sufficient to contain the cores of galaxy 
subclusters, the most critical case being the bimodal cluster A1240, 
strongly elongated in the plane of the sky.

As for the mass estimate, in a relaxed cluster, one can use the
measured global value of velocity dispersion, $\sigma_{\rm v,tot}$ to
determine $R_{\rm200}$ and then $M_{\rm 200}$ within $R_{\rm200}$.  The
estimation of mass is based on the virial theorem and, in particular,
we follow the prescriptions of Girardi et al. (\cite{gir98}; see
also Girardi \& Mezzetti \cite{gir01}). We prompt to the original
papers for the details, but, in practice, one can use:

\begin{equation}
M_{200} = A \times [\sigma_{\rm v}/(10^3 {\rm km\ s})]^3 \times
h_{70}^{-1}10^{15}{\rm M}_{\odot},
\end{equation}

\noindent where $A\sim1.4$ (here median $A=1.416$). Thus both 
our estimates of $R_{200}$ and $M_{200}$ depend on the estimate of 
the observable value of $\sigma_v$. However, they do not are mutually 
consistent according to the definition: $M_{200} = 200 \times 
\rho_{\rm crit} \times 4/3 \times \pi \times R_{200}^3$, where 
$\rho_{\rm crit}$ is the critical density of the Universe at the 
cluster redshift. Rather, according to the definition, they correspond 
to $R_{194}$ and $M_{194}$.

As for the relaxed clusters, we estimate $\sigma_{\rm v}$ from the
observed $T_{\rm X}$ assuming the equipartition of energy density
between ICM and galaxies, i.e. $\beta_{\rm spec} =1$, where
$\beta_{\rm spec}=\sigma_{\rm V}^2/(kT/\mu m_{\rm p})$ with $\mu=0.58$
the mean molecular weight and $m_{\rm p}$ the proton mass. This
assumption is particularly appropriate for massive clusters (e.g.,
Girardi et al. \cite{gir96}, \cite{gir98}). $R_{200}$ and $M_{200}$
are thus recovered with the above equations.

For each relaxed cluster, Table~\ref{tabrelax} lists the main
properties: the cluster center (Col.~2); the cluster redshift, $z$
(Col.~3); the X-ray temperature, $T_{\rm X}$ (Col.~4); the estimated
value of $\sigma_{\rm v}$ from $T_{\rm X}$ (Col.~5;) the values of
$R_{\rm 200}$ and $M_{\rm 200}$ (Cols.~6 and 7); useful references
(Col.~8).


\begin{table*}[!ht]
        \caption[]{Unrelaxed clusters (DARC sample).}
         \label{tabdarc}
              $$ 
           \begin{array}{l c c r l  c c r c}
            \hline
            \noalign{\smallskip}
            \hline
            \noalign{\smallskip}

\mathrm{Name} & \alpha,\delta\,(\mathrm{J}2000)& z &T_{\rm X}&\sigma_{\rm v,tot}& \sigma_{\rm v,subs} & M_{200} & \sigma_{\rm v}& R_{\rm 200}\\
 & & & \mathrm{keV} &\mathrm{\,km}\mathrm{s^{-1}\,}&\mathrm{(\,km}\mathrm{s^{-1}\,)}&10^{15} {\rm M}_{\sun}&\mathrm{(\,km}\mathrm{s^{-1}\,)}&\mathrm{Mpc}\\
            \hline
            \noalign{\smallskip}  
{\rm A697}^{\mathrm{a}}  & 08\ 42\ 57.55 , +36\ 21\ 59.9& 0.2815& 10.2& 1334_{-95}^{+114}&200,600,400&0.40&660& 1.39\\
{\rm A773}^{\mathrm{b}} & 09\ 17\ 53.26 , +51\ 43\ 36.5& 0.2197&  7.8& 1394_{-68}^{+84} &950,500&1.39&994&2.16\\
{\rm A959}^{\mathrm{c}}  & 10\ 17\ 35.04 , +59\ 33\ 27.7& 0.2883&  7.0& 1170_{-73}^{+83} &600,700&0.76&824&1.73\\
{\rm A1240}^{\mathrm{d}} & 11\ 23\ 37.60 , +43\ 05\ 51.0& 0.1948&  6.0&  870_{-79}^{+91} &700,1000&1.93&1103& 2.43\\
{\rm A2219}^{\mathrm{e}} & 16\ 40\ 19.87 , +46\ 42\ 41.3& 0.2254& 10.3& 1438_{-86}^{+109}&1000&1.41&1000&2.17\\
                        \noalign{\smallskip}			    
            \hline					    
            \noalign{\smallskip}			    
            \hline					    
         \end{array}
     $$ 
\begin{list}{}{}

\item[$^{\mathrm{a}}$] Ref.: Girardi et al. \cite{gir06};  
$\sigma_{\rm v,subs}$ from KMM4g1-2-4 in Table~4.
\item[$^{\mathrm{b}}$] Ref.: Barrena et al. \cite{bar07}; $\sigma_{\rm v,subs}$ from the main and secondary subclusters in Sect.~5.
\item[$^{\mathrm{c}}$] Ref.: Boschin et al. \cite{bos08}; $\sigma_{\rm v,subs}$ from the averages of
  KMM1 and V1, and of KMM2 and V2 in Table~2. Here we not consider the
  subclump V3(=KMM3=DS-NE) also detected in X-ray and thus likely in a
  premerging phase and not responsable of the ongoing merger.
\item[$^{\mathrm{d}}$] Ref.: Barrena et al. \cite{bar09}; $\sigma_{\rm v,subs}$ from
A1240N and A1240S in Table~2.
\item[$^{\mathrm{e}}$] Ref.: Boschin et al. \cite{bos04}; $\sigma_{\rm
  v,subs}$ from NW in Table~2. The separation of suclumps is not
  obvious. Here, since the primary subclump lies at the SE of the cluster
  center (see also the more recent paper Million \& Allen
  \cite{mil09}), we consider the value of $\sigma_{\rm v}$ of NW
  sector, where the main cluster is likely free from the subclump.

\end{list}
\end{table*}


As for the DARC clusters we have to adopt a more complex procedure. 
During a cluster merger, the global value of velocity dispersion, 
$\sigma_{\rm v,tot}$, might strongly vary and not be a very good 
indicators of the real cluster potential (e.g., Pinkney et al. 
\cite{pin96}) and the same problem could afflict $T_{\rm X}$ (e.g., 
Ricker \& Sarazin \cite{ric01}; Mastropietro \& Burkert \cite{mas08}). 
As discussed in our papers for the five DARC clusters, 
we have detected the main subclusters and obtain an estimate of their 
individual velocity dispersion, $\sigma_{\rm v,subs}$. Thus, an 
alternative, likely more reliable value of the mass $M_{200}$ of the 
cluster, can then be obtained by adding the masses of all subclusters.  
Then, inverting the above scaling relation eq.~2, we can obtain a 
value of the velocity dispersion, $\sigma_{\rm v}$, which should be 
a better indicator of the potential. The use of eq.~1 leads to the 
estimate of $R_{200}$.

For each DARC cluster, Table~\ref{tabdarc} lists the main
properties: the cluster center (Col.~2); the cluster redshift, $z$
(Col.~3); the X-ray temperature, $T_{\rm X}$ (Col.~4); the global
observed value of velocity dispersion, $\sigma_{\rm v,obs}$ (Col.~5) ;
the (rounded) values of velocity dispersions for subclusters
$\sigma_{\rm v,subs}$ (Col.~6); the estimated value of mass from the
addition of subclusters, $M_{200}$ (Col.~7); the corresponding values
of $\sigma_{\rm v}$ and $R_{\rm 200}$ (Cols.~8 and 9). The values of
the cluster center, $z$, $T_{\rm X}$, $\sigma_{\rm v,tot}$, and
$\sigma_{\rm v,subs}$ can be obtained from the listed references. In
particular, the notes list the source for the values of
$\sigma_{\rm v,subs}$, here rounded. Other values are homogeneously
computed in this study.


\begin{table*}[!ht]
        \caption[]{Relaxed cluster (Comparison sample).}
         \label{tabrelax}
              $$ 
           \begin{array}{l c c r  r c c l}
            \hline
            \noalign{\smallskip}
            \hline
            \noalign{\smallskip}

\mathrm{Name} & \alpha,\delta\,(\mathrm{J}2000)&z&T_{\rm X}&\sigma_{\rm v}& R_{\rm 200}&M_{200}&{\rm Refs.}^{\mathrm{a}}\\
 & & & \mathrm{keV} &\mathrm{(\,km}\mathrm{s^{-1}\,)}& \mathrm{Mpc}& 10^{15} {\rm M}_{\odot}&\\
            \hline
            \noalign{\smallskip}  
{\rm A383}  & 02\ 48\ 03.50 , -03\ 31\ 45.0 & 0.188& 3.9&802&1.78&0.74&{\rm Allen\ et\ al.\ \cite{all04}, Maughan\ et\ al.\ \cite{mau08}}\\
{\rm A963}  & 10\ 17\ 01.20 , +39\ 01\ 44.4 & 0.206& 6.0&995&2.18&1.41&{\rm Allen\ et\ al.\ \cite{all04}, Baldi\ et\ al.\ \cite{bal07}}\\
{\rm A1835} & 14\ 01\ 02.40 , +02\ 52\ 55.2 & 0.252& 8.1&1156&2.47&2.15&{\rm Allen\ et\ al.\ \cite{all04},Baldi\ et\ al.\ \cite{bal07}}\\
{\rm A2261} & 17\ 22\ 27.60 , +32\ 07\ 57.2 & 0.224& 7.2&1090&2.36&1.83&{\rm Baldi\ et\ al.\ \cite{bal07},Maughan\ et\ al.\ \cite{mau08}}\\
{\rm ZwCl3146} & 10\ 23\ 39.60 , +04\ 11\ 24.0& 0.291& 8.6&1191&2.49&2.31&{\rm Baldi\ et\ al.\ \cite{bal07},Baldi\ et\ al.\ \cite{bal07}}\\
                        \noalign{\smallskip}			    
            \hline					    
            \noalign{\smallskip}			    
            \hline					    
         \end{array}
$$
\begin{list}{}{}
\item[$^{\mathrm{a}}$] References for $z$
  and $T_{\rm X}$ are listed in the first and second
  position, respectively. Cluster centers are taken from Ebeling et
  al. \cite{ebe96} or Ebeling et al. \cite{ebe98}.
\end{list}
\end{table*}


\section{Magnitude data and background contamination}
\label{data}

The magnitude data ($r'$,$i'$) are taken from SDSS (Data Release
7). We retrieve {\it model mag} values. For each cluster, we 
create a galaxy ``cluster galaxy sample'' by selecting the galaxies
in an area centered on the cluster center listed in
Tables~\ref{tabrelax} and \ref{tabdarc} and within the respective
radius $R_{200}$. We verify that the whole cluster area is covered
by SDSS data by visual inspection.  

\begin{figure}
\centering 
\resizebox{\hsize}{!}{\includegraphics{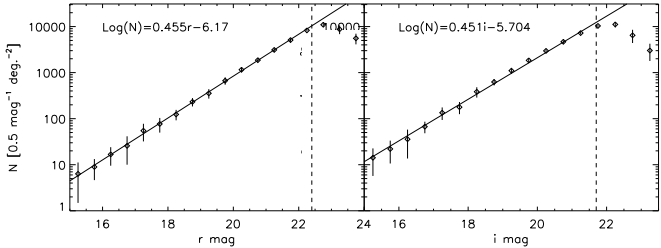}}
\caption{Values of $r'-$ and $i'-$band counts for the 10 square degrees
  considered as the field galaxy sample. The solid line
  indicates the logarithmic function fit to the histograms. The dashed
  line indicates the completeness magnitudes.}
\label{fig:field_ri}
\end{figure}

For each cluster, we also consider a galaxy sample selecting galaxies
in one square degree field at 1.5 degree toward the west of each
cluster obtaining ten individual field samples.  Together these
samples, for a total area of ten square degrees, forms the ``field
galaxy sample'' and are used to estimate the field galaxy
contamination. Fig.~\ref{fig:field_ri} shows the field counts for $r'$
and $i' $ magnitudes, and the logarithmic function fit. This fit
represents the count-magnitude relation expected for a homogeneous
galaxy distribution in a universe with Euclidean geometry.  We obtain
the relations $\log(N_r)=0.455(\pm0.043)r'-(6.17\pm0.59) \ 0.5mag^{-1}
deg^{-2}$ for $15<r'<22$ and
$\log(N_i)=0.451(\pm0.046)i'-(5.70\pm0.58) \ 0.5mag^{-1} deg^{-2}$ for
$15<i'<21.5$. We used these fits with two purposes: i) to estimate the
completeness of our sample; ii) to subtract field galaxy contamination
in each cluster.

By comparison with the logarithmic fits, the completeness magnitude can 
be derived. So, we estimate the photometric sample is complete down to
the magnitude counts are lower than 5\% of the fit. Therefore, we 
consider the $r'$- and $i'$-band photometry complete down to $r'=22.4$ 
and $i'=21.4$, respectively. We find that these values are in very 
good agreement with those reported by the Sloan Digital Sky Survey DR7 
(see \verb+ http://www.sdss.org/dr7/+).

Two different approaches to the statistical subtraction of the galaxy
foreground/background are discussed in the literature: several studies
have examined the effect on the derived cluster LF using a global
galaxy field correction versus one measured locally for each
cluster. For instance, Popesso et al. (\cite{pop04}) has shown, for a
study of about 100 clusters based on SDSS data, that there is no
significant difference in the measured cluster LF parameters using
either a global or local field subtraction technique. Here we consider
the global field correction, subtracting the field counts derived
from the logarithmic fits to the cluster counts. Note that the field
counts are treated in a different and appropriate way for each cluster
before the subtraction.

The magnitudes of each cluster are  corrected
for the galactic absorption using the Schlegel, Finkbeiner and Davis 
Galactic reddening maps (Schlegel et al. \cite{Sch98}), derived from
IRAS and COBE/DIRBE data.

\begin{figure}
\centering 
\resizebox{\hsize}{!}{\includegraphics{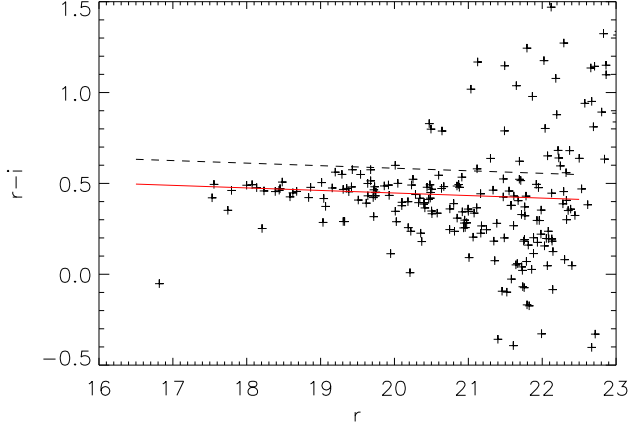}}
\caption{Color magnitude diagram $r'-i'$ vs. $r'$ considering 
galaxies within $0.5$ \h from the center of Abell 383. 
The RS (red solid line) is fitted using  galaxies with 
$r'<20.2$. Galaxies with $r'-i'>$RS+0.15 (on the dashed line) 
are considered background galaxies and  rejected from the analysis.}
\label{fig:cmr_a383}
\end{figure}

\begin{figure}
\centering 
\resizebox{\hsize}{8.8cm}{\includegraphics{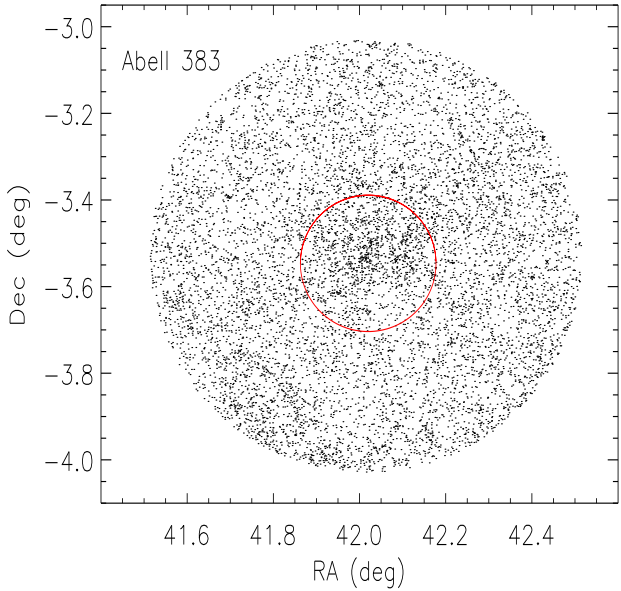}}
\caption{Galaxies with $r'-i'<$RS+0.15 in the region of A383. The red
circle indicates $R_{200}$ for this cluster.}
\label{fig:clustfield}
\end{figure}

The possible contamination of distant clusters projected onto the
field-of-view of the target cluster can prove to be problematic in the
derivation of the LF, in particular at the faint-end. One can use that
the spectroscopic studies show that there are essentially no cluster
galaxies significantly redward of the red sequence (e.g. Rines \&
Geller \cite{rin08}). We minimize the contamination of background
clusters using the $r'-i'$ vs.  $r'$ colour-magnitude relation which
is proved to be the most useful from target clusters are at
$0.2<z<0.4$ (Lu et al. \cite{lu09}, their Fig.~6 and Sect.~3.4.1). For
each cluster, we fit the red sequence (RS) $r'-i'$ vs. $r'$ for
galaxies. We reject galaxies that are 0.15 mag redward of the RS
(i.e. $\gtrsim 3$ times the average dispersion of the cluster RS).
Fig.\ref{fig:cmr_a383} shows an example for the RS in the case of
A383. After this background correction, the clusters appear more
contrasted with respect to the surrounding field (see
Fig.~\ref{fig:clustfield} for A383.)

Tab. \ref{tab:cmr} lists, for each cluster, the slope $a$ and the
intercept $b$, and standard deviations, of the RS $r'-i'$ vs. $r'$
obtained with a linear fit within $0.5$ \h from the center and for
galaxies with $r'<20.2$ (Col.~2 and 3).


\begin{table}[!ht]
        \caption[]{Color-magnitude relations (CMR) $r'-i'$ vs. $r'$ linear 
	fits ($a \dot z + b$): Red Sequences.}
         \label{tab:cmr}
              $$ 
           \begin{array}{l  c c}
            \hline
            \noalign{\smallskip}
            \hline
            \noalign{\smallskip}

\mathrm{Name} & \mathrm{(a,\delta a)} & \mathrm{(b,\delta b)}       \\
            \hline
            \noalign{\smallskip}  

A697  &  -0.0107,0.014 & 0.744,0.041 \\
A773  &  -0.0213,0.007 & 0.861,0.037 \\
A959  &  -0.0310,0.006 & 1.126,0.037 \\
A1240 &  -0.0110,0.020 & 0.666,0.037 \\
A2219 &  -0.0086,0.007 & 0.668,0.062 \\
\hline
\noalign{\smallskip}  
A383     & -0.0160,0.001 & 0.763,0.047 \\
A963     & -0.0422,0.007 & 1.216,0.042 \\
A1835    & -0.0004,0.011 & 0.501,0.031 \\
A2261    & -0.0091,0.015 & 0.653,0.048 \\
ZwCl3146 & -0.0220,0.003 & 1.020,0.050 \\		       
            \noalign{\smallskip}			    
            \hline					    
            \noalign{\smallskip}			    
            \hline					    
         \end{array}
     $$ 
\end{table}

\begin{figure}
\centering 
\resizebox{\hsize}{!}{\includegraphics{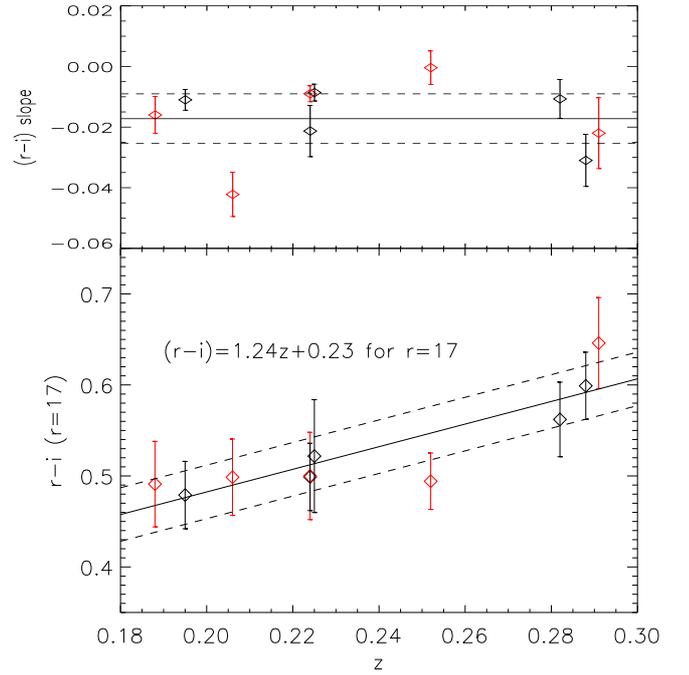}}
\caption{RS parameters vs. redshift: the $(r'-i')$ vs. $r'$ slope ({\em
    upper panel}) and $(r'-i')$ color at $r'=17$ ({\em lower panel}).
  Black symbols indicate DARC clusters and red symbols indicate
  relaxed clusters.  The solid line in the upper panel shows the
  mean value for all ten clusters.  The solid line in the lower
  panel shows the linear fit.  The dashed lines show one-sigma
  uncertainties of the linear fit.}
\label{fig:ri_z}
\end{figure}

Fig.~\ref{fig:ri_z} shows the RS parameters vs. cluster redshift.
As for the RS slope, there is no correlation with $z$ in the observed
range and we estimate a mean value of $<a>=-0.017 \pm 0.008$.  As for
the $r'-i'$ colors at $r'=17$, as computed from the RS fits, there is
no difference in the behaviour of DARC and relaxed clusters: we only
observe a marginally higher value for the most distant cluster A697,
A959 and ZwCl3146, which is consistent with the $k-$correction effect 
on early-type galaxies in the central core of clusters (see
Fukugita et al. (\cite{fuk95}, tables 3, 6, 7 and 8). Assuming a linear 
variation of the color with the redshift, we obtain the relation
$(r'-i')_{r'=17}=1.24*z+0.23$, with a global dispersion in color of
$\pm0.02$.

For each cluster, we transform galaxy apparent into absolute magnitudes by
applying the relation:

\begin{equation}
M_r=r'-2.5-5{\rm log}(D_L/1{\rm Mpc})-K(z)+E(z),
\end{equation} 

\noindent where $r'$ is the apparent magnitude (already corrected for
the Galactic absorption, see above), $D_L$ is the luminosity distance;
$K(z)$ is the K-correction and $E(z)$ is the
evolutionary correction. $K(z)$ and $E(z)$ corrections are applied using a
single parametrization based on early--type galaxies, which are
dominant in the cluster galaxy population. As for $K(z)$, we use the
correction obtained from the Table~2, col. 3 in Roche et al. (\cite{roc09}). 
As for the evolutionary correction, we use $E(z) = 0.86z$
(Roche et al. \cite{roc09}). Given that the completeness of magnitude
data is $r'=22.4$ (Fig.~\ref{fig:field_ri}) and the most distant
cluster of our catalogs is at $z=0.29$, we expect that the LFs we
obtain in the present study are all reliable for absolute magnitude
$M_r<-18.6$.

For each cluster, before the field subtraction in the construction of
the LF (see below), the field galaxies are treated in the same way of
the respective cluster galaxies, i.e. we apply the same correction for
the Galactic absorption; the same rejection of very red galaxies; the
same transformation in absolute magnitudes.

\section{Galaxy populations in DARC and relaxed clusters}
\label{lf}

\subsection{Individual luminosity functions}
\label{sub:lf}

\begin{figure}
\centering 
\resizebox{\hsize}{!}{\includegraphics{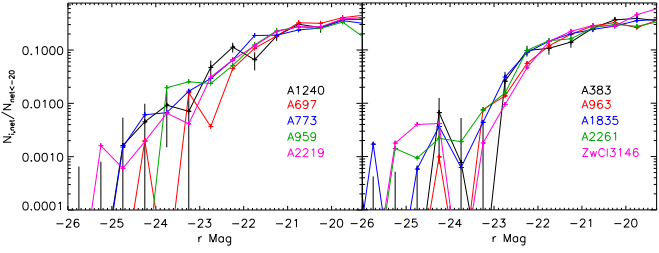}}
\caption{LFs for DARC clusters ({\em right panel}) and relaxed clusters
({\em left panel}) as
  computed within $R_{200}$. Error bars are only plotted in the
  A1240 and A383 LFs for the sake of clarity.}
\label{fig:fdl_rvir}
\end{figure}

Fig.~\ref{fig:fdl_rvir} shows the individual LFs for DARC and
relaxed clusters within $R_{200}$, where the galaxy counts are
computed as $N_i=N_{net,i}/N_{{\rm net},<-20}$.  $N_{net,i}$ are the
net counts, i.e. $N_{net,i}=N_{\rm{cluster},i}-N_{{\rm field},i}$
where the field counts are opportunely normalized to the cluster
area. $N_{{\rm net},<-20}$ is the number of net counts for $M_r<-20$
and thus represents the ``richness'' here used to normalize.  The
error bars in Fig.~\ref{fig:fdl_rvir} are the richness-normalized
$\sigma_i=(N_i+N_{b,i}+rms_f^2)^{1/2}$, where the first two terms take
into accounts the Poissonian errors of the cluster and background
field samples, and the third term measures the field-to-field
variation per bin in the background field counts (i.e., the cosmic
variance obtained using the ten individual field samples). In order
to see possible variations with the radius, we also compute the LFs 
within $0.5R_{200}$ and $R_{200}$.

Each LF was fitted to a Schechter function (see Schechter
\cite{sch76}) using the ``minuit'' procedure (CERN Libraries) to
minimize the three values, $\phi$ (richness), $\alpha$ (slope) and
$M^*$ of the Schechter profile. We do not discuss the $\phi$
parameter since here the counts are already normalized by the
``richness'' $N_{net,<-20}$ for each cluster.
Table~\ref{tab:alfam} shows $M^*$ and $\alpha$ values obtained for each
cluster, fitting Schechter function within $R_{200}$ and
$0.5R_{200}$. 


\begin{table}[!ht]
        \caption[]{$M_r^\star$ and $\alpha$ values obtained from Schechter
	fit for the LF.}
         \label{tab:alfam}
              $$ 
           \begin{array}{l c c c c}
            \hline
            \noalign{\smallskip}
            \hline
            \noalign{\smallskip}
              & \multicolumn{2}{c}{R<R_{200}} & \multicolumn{2}{c}{R<0.5R_{200}} \\
\mathrm{Name} & M_r^\star, \Delta M_r^\star & \alpha, \Delta \alpha & 
                M_r^\star, \Delta M_r^\star & \alpha, \Delta \alpha \\
           \hline
            \noalign{\smallskip}  

A697  & -21.19,0.30 & -0.96,0.20 & -20.95,0.16 & -0.70,0.21 \\
A773  & -21.84,0.26 & -1.01,0.13 & -21.43,0.24 & -0.61,0.18 \\
A959  & -21.60,0.62 & -0.96,0.39 & -22.00,0.94 & -0.96,0.46 \\
A1240 & -22.31,0.36 & -1.28,0.12 & -21.61,0.37 & -0.95,0.20 \\
A2219 & -21.74,0.28 & -1.09,0.15 & -21.79,0.08 & -0.97,0.08 \\
\hline
            \noalign{\smallskip}  
A383     & -21.89,0.25 & -1.30,0.17 & -21.92,0.27 & -1.26,0.21 \\
A963     & -21.31,0.28 & -0.83,0.19 & -20.93,0.22 & -0.27,0.25 \\
A1835    & -21.78,0.21 & -1.02,0.15 & -22.06,0.33 & -1.13,0.18 \\
A2261    & -21.56,0.22 & -0.88,0.16 & -21.62,0.25 & -1.01,0.19 \\
ZwCl3146 & -21.36,0.26 & -0.97,0.24 & -20.89,0.23 & -0.51,0.29 \\			
            \noalign{\smallskip}			    
            \hline					    
            \noalign{\smallskip}			    
            \hline					    
         \end{array}
     $$ 
\end{table}

\begin{figure}
\centering 
\resizebox{\hsize}{!}{\includegraphics{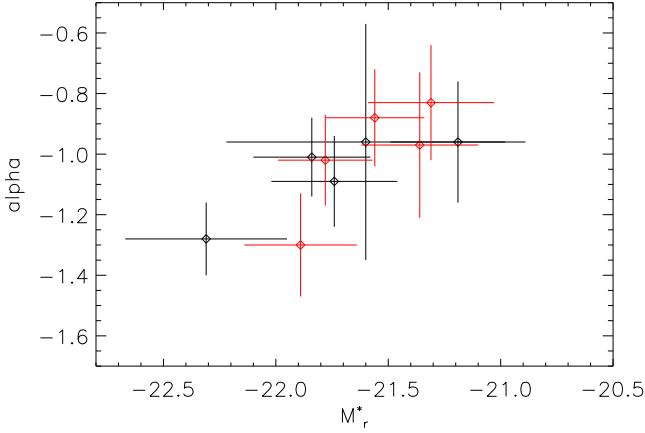}}
\caption{$\alpha$ vs. $M_r^\star$ derived from the Schechter LF fit. 
Black symbols indicate DARC clusters and red symbols indicate relaxed 
clusters.}
\label{fig:mstar_alpha}
\end{figure}

Fig.~\ref{fig:mstar_alpha} shows $\alpha$ vs. $M^*$ values estimated 
for each cluster within $R_{200}$: the absence of difference between 
DARC and relaxed clusters is shown. We obtain the same result for LFs 
within $0.5R_{200}$ and when comparing $R_{200}$ and $0.5R_{200}$ results.

For the DARC sample we obtain a mean value of $M^*_{\rm DARC}=-21.73
\pm 0.16$ and $-21.57 \pm 0.16$, considering galaxies within $R_{200}$
and $0.5R_{200}$, respectively. On the other hand, for the relaxed
clusters, we derive $M^*_{\rm rel}=-21.58 \pm 0.11$ and $-21.48 \pm
0.12$. For the slope parameter, we estimate $\alpha_{\rm DARC}=-1.06
\pm 0.09$ and $-0.84 \pm 0.10$ within $R_{200}$ and $0.5R_{200}$,
respectively.  For the relaxed clusters, we obtain $\alpha_{\rm
rel}=-1.00 \pm 0.12$ and $-0.84 \pm 0.10$. Due to the correlated 
nature of $\alpha$ and $M^\star$ parameters in the fitting
procedure, which likely leads to the apparent correlation in Fig.
~\ref{fig:mstar_alpha}, we prefer to compare individual clusters 
fixing one parameter and remake the fit. When fixing $\alpha=-1.00$ 
for the slope parameter for all individual clusters, we find no longer 
difference (see Table~\ref{tab:mstaralfa1}). In particular, when fixing 
the slope, we obtain $M^*_{DARC}=-21.64 \pm 0.06$ and $M^*_{rel}=-21.61 
\pm 0.05$ within $R<R_{200}$ and $M^*_{DARC}=-21.82 \pm 0.14$ and
$M^*_{rel}=-21.64 \pm 0.06$ within $R<0.5R_{200}$. From this
results, we can conclude that the global LF profile is similar for
DARC and relaxed clusters and independent of the sampling radius.


\begin{table}[!ht]
        \caption[]{$M_r^\star$ values obtained from the Schechter
	fit when fixing $\alpha=-1.00$ for the LF.}
         \label{tab:mstaralfa1}
              $$ 
           \begin{array}{l c c}
            \hline
            \noalign{\smallskip}
            \hline
            \noalign{\smallskip}
              & R<R_{200} & R<0.5R_{200} \\
\mathrm{Name} & M_r^\star, \Delta M_r^\star & M_r^\star, \Delta M_r^\star \\
           \hline
            \noalign{\smallskip}  

A697  & -21.30,0.12 & -21.19,0.79 \\
A773  & -21.85,0.11 & -22.15,0.12 \\
A959  & -21.83,0.16 & -22.25,0.19 \\
A1240 & -21.65,0.17 & -21.69,0.18 \\
A2219 & -21.57,0.12 & -21.81,0.31 \\
\hline
            \noalign{\smallskip}  
A383     & -21.51,0.14 & -21.65,0.21 \\
A963     & -21.54,0.12 & -21.70,0.13 \\
A1835    & -21.75,0.09 & -21.95,0.14 \\
A2261    & -21.72,0.09 & -21.59,0.11 \\
ZwCl3146 & -21.53,0.13 & -21.33,0.08 \\ 		      
            \noalign{\smallskip}			    
            \hline					    
            \noalign{\smallskip}			    
            \hline					    
         \end{array}
     $$ 
\end{table}


The mean LF will be studied in the following subsection, and we will 
detail possible differences considering blue and red galaxy population
in the inner and external regions.

\subsection{The composite luminosity functions}

For both DARC and relaxed clusters separately, we construct the
composite LFs by combining into a single "ensemble" LF, in order to
improve the rather poor number statistics for each individual LF. The
composite LF is built by averaging, in absolute magnitude bins, the
richness-normalized net counts as obtained above for each individual
LF. We assume the standard errors on the average. 

\begin{figure}
\centering 
\resizebox{\hsize}{!}{\includegraphics{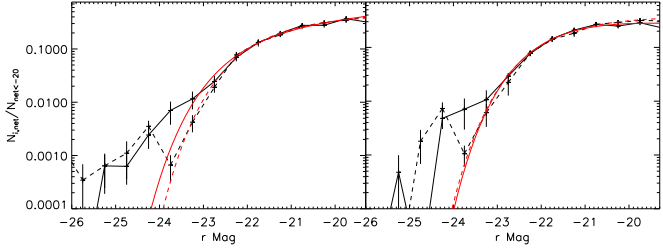}}
\caption{Composite LFs within $R_{200}$ ({\em left panel}) and $0.5R_{200}$
({\em right panel}). Solid lines show the counts for DARC clusters and
dashed lines show the counts for the relaxed clusters. The red lines
indicate the corresponding Schechter fits.}
\label{fig:fdl_tot}
\end{figure}

Fig.~\ref{fig:fdl_tot} shows the composite LFs for DARC and relaxed
clusters within $R_{200}$ and $0.5R_{200}$. Table~ \ref{tab:sctot}
summarizes the $M^*$ and $\alpha$ parameters obtained for the
composite LFs. In agreement with the above section, there is no
evidence of difference between DARC and relaxed samples and there is
no dependence on the sampling radius.

\begin{table}[!ht]
        \caption[]{$M^*$ and $\alpha$ for the composite LFs.}
         \label{tab:sctot}
              $$ 
           \begin{array}{l c c c c}
            \hline
            \noalign{\smallskip}
            \hline
            \noalign{\smallskip}
                & \multicolumn{2}{c}{R<R_{200}} & \multicolumn{2}{c}{R<0.5R_{200}}  \\
            \hline
            \noalign{\smallskip}
\multicolumn{5}{c}{\textrm{Whole galaxy population}} \\
\mathrm{Sample} & M^*,\delta M^* & \alpha,\Delta\alpha & M^*,\delta M^* & \alpha,\Delta\alpha \\
            \hline
            \noalign{\smallskip}  
DARC    & -22.00, 0.41 & -1.20, 0.17 & -21.64, 0.35 & -0.86, 0.27 \\
Relaxed & -21.63, 0.24 & -1.03, 0.20 & -21.74, 0.29 & -0.99, 0.21 \\
            \noalign{\smallskip}			   
\multicolumn{5}{c}{\textrm{Red galaxies}} \\
            \hline
            \noalign{\smallskip}			   
DARC    & -21.62, 0.25 & -0.76, 0.22 & -21.23, 0.28 & -0.47, 0.27 \\
Relaxed & -21.51, 0.28 & -0.76, 0.23 & -21.50, 0.36 & -0.70, 0.30 \\
            \noalign{\smallskip}			   
\multicolumn{5}{c}{\textrm{Blue galaxies}} \\
            \hline
            \noalign{\smallskip}			   
DARC    & -22.55, 0.27 & -2.03, 0.10 & -22.98, 0.37 & -1.79, 0.24 \\
Relaxed & -21.66, 0.52 & -1.44, 0.46 & -23.63, 0.53 & -2.07, 0.23 \\
            \hline					    
            \noalign{\smallskip}			    
            \hline					    
         \end{array}
     $$ 
\end{table}



Fig.~\ref{fig:fdl_tot} shows as the Schechter functions fit reasonably
well the counts for galaxies with $M_{r'}>-22.5$. We obtain
\footnote{The $\chi^2$ statistic has been calculated considering three
degrees of freedom, the normalization coefficient, $M^\star$ and the
slope.} $\chi^2=0.45$ for DARC clusters and 0.35 for relaxed
clusters (both within $R_{200}$) and $\chi^2=0.33$ for DARC clusters
and 0.27 for relaxed clusters (both within $0.5 R_{200}$). On the
contrary, the bright-end ($M_{r'}<-22.5$) is not well represented by
Schechter functions. In any case, we obtain $\chi^2$ values higher than
1.4. Therefore, the contribution of these bright galaxies (hereafter
BGs) have to be analyzed separately (see Sect.~\ref{bg}).

\subsection{Blue and red LFs}

\begin{figure}
\centering 
\resizebox{\hsize}{!}{\includegraphics{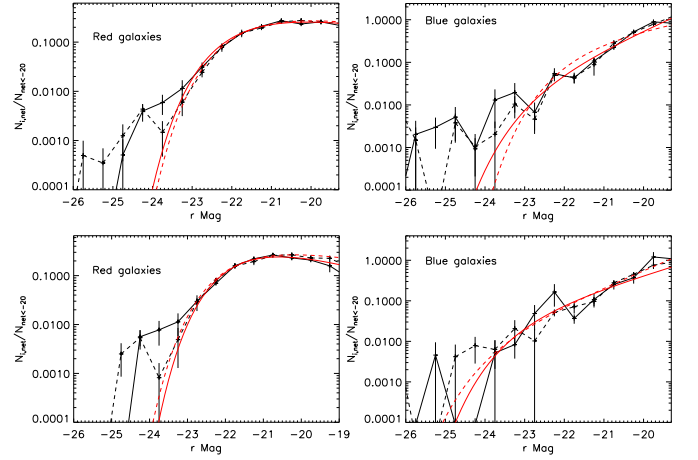}}
\caption{{\em Upper panels}: LFs and Schechter fits for red and blue galaxies
  computed within $R_{200}$. {\em Lower panels}: the same within
  $0.5R_{200}$. Solid lines indicate DARC clusters and 
dashed lines indicate relaxed clusters.}
\label{fig:fdl_blue_red}
\end{figure}

We also estimate LF for red and blue galaxies within $R_{200}$ and
$0.5R_{200}$. We select red and blue galaxies from the RS (see
above). We assume red galaxies those objects within the RS, that is
RS-0.15$<r'-i'<$RS+0.15, while blue galaxies were selected as
$r'-i'<$RS-0.15. Results are presented if Fig.~\ref{fig:fdl_blue_red}
and summarized in Table~\ref{tab:sctot}.

\begin{figure}
\centering 
\resizebox{\hsize}{!}{\includegraphics{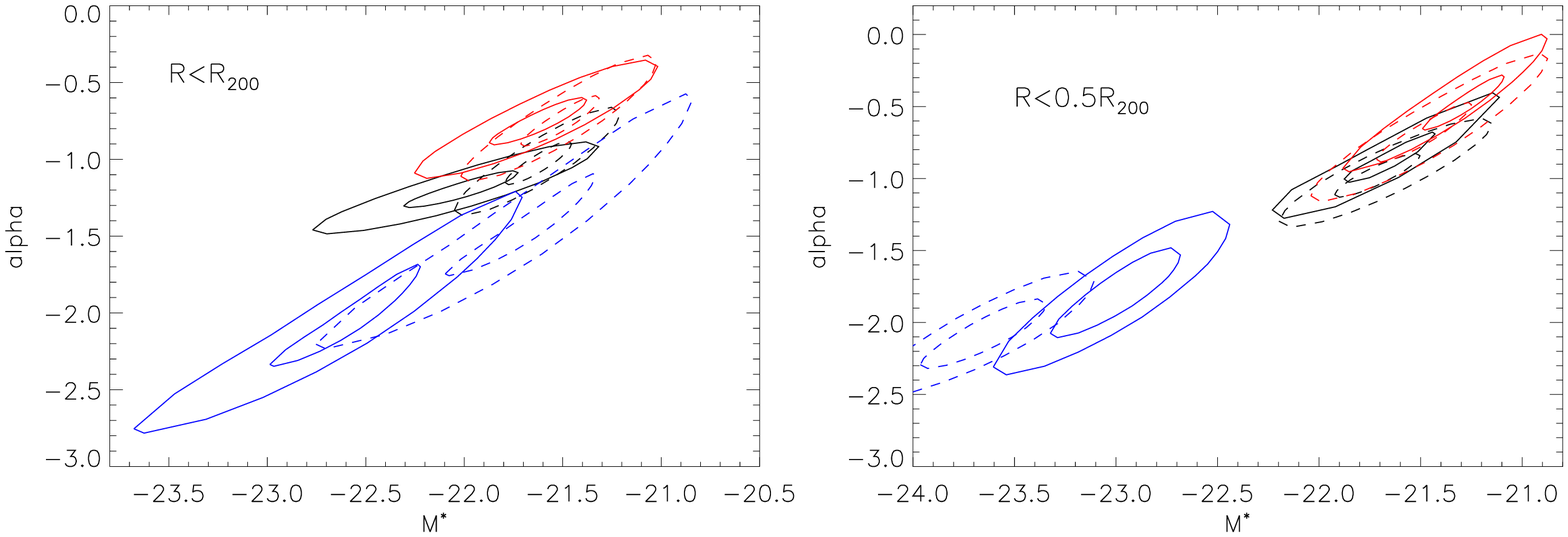}}
\caption{Error contours at the 1 and 3$\sigma$ c.l. for the
  best-fitting LF parameters considering the galaxy population within
  $R<R_{200}$ ({\em left panel}) and $R<0.5R_{200}$ ({\em right
    panel}). Black, red and blue lines show the composite LF for the
  whole galaxy sample, red and blue galaxies, respectively.  Solid
  lines indicate DARC clusters and dashed lines indicate relaxed
  clusters.}
\label{fig:ell_blue_red}
\end{figure}

When considering the whole galaxy population, no differences are 
found between DARC and relaxed clusters within $R_{200}$ or inside the 
core of the clusters, within $0.5R_{200}$. The differences in $M^\star$ 
and $\alpha$ are not significant and within 3$\sigma$ errors (see 
Fig.~\ref{fig:ell_blue_red}). The same is also true for the red galaxy 
population. We only note a marginal difference in the LF parameters 
of blue galaxies, in particular in the core of the clusters, inside 
$0.5R_{200}$. 

To check the significance of these differences, we fit the Schechter
profiles to the composite LFs fixing the slope
parameter. Table~\ref{tab:sctot_a1} lists the results obtained. Again,
we confirm no differences between DARC and relaxed LFs. As above, we
only find small differences for the blue galaxy population within
$0.5R_{200}$. However, this difference should be taken with care
since the blue galaxy population is very poor in the internal regions
of the clusters thus limiting number of considered counts.


\begin{table}[!ht]
        \caption[]{$M^*$ for the composite LFs (at fixed $\alpha$).}
         \label{tab:sctot_a1}
              $$ 
           \begin{array}{l c c }
            \hline
            \noalign{\smallskip}
            \hline
            \noalign{\smallskip}
                & R<R_{200} & R<0.5R_{200}  \\
            \hline
            \noalign{\smallskip}
\multicolumn{3}{c}{\textrm{Whole galaxy population} \ \ (\alpha=-1.00)} \\
\mathrm{Sample} & M^*,\delta M^* & M^*,\delta M^* \\
            \hline
            \noalign{\smallskip}  
DARC    & -21.61, 0.07 & -21.83, 0.06 \\
Relaxed & -21.59, 0.05 & -21.75, 0.06 \\
            \noalign{\smallskip}			   
\multicolumn{3}{c}{\textrm{Red galaxies} \ \ (\alpha=-0.70)} \\
            \hline
            \noalign{\smallskip}			   
DARC    & -21.54, 0.07 & -21.52, 0.05 \\
Relaxed & -21.44, 0.05 & -21.48, 0.06 \\
            \noalign{\smallskip}			   
\multicolumn{3}{c}{\textrm{Blue galaxies} \ \ (\alpha=-2.00)} \\
            \hline
            \noalign{\smallskip}			   
DARC    & -22.74, 0.52 & -23.18, 0.11 \\
Relaxed & -22.38, 0.11 & -23.56, 0.06 \\
            \hline					    
            \noalign{\smallskip}			    
            \hline					    
         \end{array}
     $$ 
\end{table}


\subsection{Butcher-Oemler effect}
\label{bri}

The window opened by the redshift dependence of the galaxy properties 
has been used to set constraints on the time-scales of the processes
of galaxy evolution (Butcher \& Oemler, \cite{but78}, \cite{but84}; 
Stanford et al. \cite{sta98}). In this context, 
the observational evidence of the environmental effect is still uncertain. 
In order to clarify whether merging processes in clusters can enhance the
star formation in the galaxy populations, we compare the fraction 
of blue galaxies, $f_{\rm B}$, in DARC and relaxed clusters. So we
quantify the Butcher-Oemler effect (Butcher \& Oemler, \cite{but78}, 
\cite{but84}) in two homogeneous cluster samples with similar velocity
dispersions and masses.

Studies of Andreon et al. (\cite{and06}) show that the radial dependence 
of the blue fraction is quite shallow, and it smoothly and monotonically 
increases from the center of the cluster to the field (see Fig. 10 in
Andreon et al. \cite{and06}). However, the contamination contribution
of field galaxies is lower in the inner regions. For this reason, we 
estimate $f_{\rm B}$ within $0.5R_{200}$.

\begin{figure}
\centering 
\resizebox{\hsize}{!}{\includegraphics{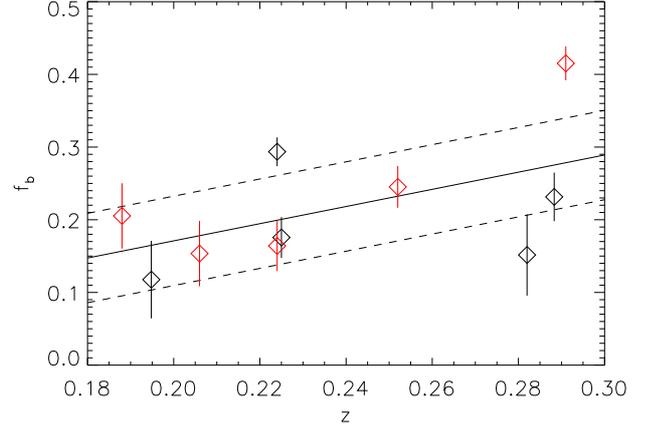}}
\caption{Fraction of blue galaxies within $0.5R_{200}$ for DARC (black symbols) 
and relaxed clusters (red symbols). The solid line describes the linear fit of
this relation. The dashed lines show one-sigma uncertainties of the 
linear fit.}
\label{fig:BO}
\end{figure}

Fig. \ref{fig:BO} shows as $f_{\rm B}$ increases towards high redshifts 
being at the 95\% c.l. according to the Kendall statistics. A
linear fit gives $f_B(z)=1.18(\pm0.07)z-0.07(\pm 0.06)$ (see also
Sect.~\ref{disc} for further discussions). We find no significant
difference between $f_{\rm B}$ of DARC and relaxed clusters.

\section{The bright end of the LF}
\label{bg}

Here we define bright galaxies (BGs) those galaxies with $M_{r'}<-22.5$ 
and analyze whether DARC and relaxed clusters differ for their BGs,
in counts or luminosity. 

\subsection{Counting BGs}
\label{cf}

\begin{figure}
\centering 
\resizebox{\hsize}{!}{\includegraphics{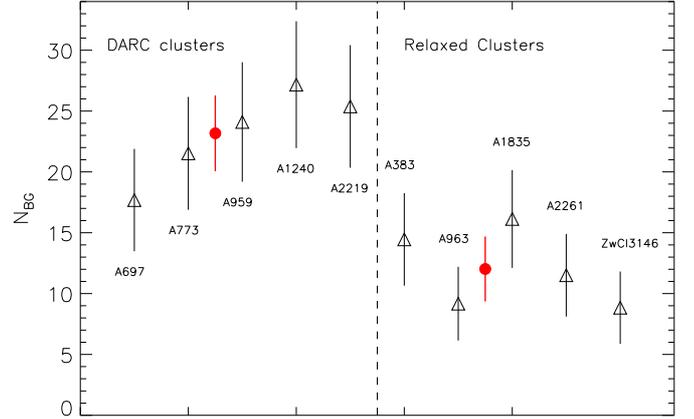}}
\caption{Number of bright galaxies within $R_{200}$. Vertical dashed 
lines separate DARC and relaxed cluster samples. Labels indicate individual
clusters. The red points represent the mean value computed for 
each cluster sample separately.}
\label{fig:BG}
\end{figure}

Fig.~\ref{fig:BG} shows the number of BGs for each cluster
$N_{\rm BG}=N_{r'<-22.5}/N_{r'<-20}\dot<N>$, where $N_{r'<-22.5}$
is the number of galaxies brighter than $M_{r'}<-22.5$ (normalized to
the cluster richness, see the above Sect.~\ref{lf}) and $<N>=<N_{r'<-20}>$ is
the mean richness as computed on all clusters. Here we multiply for
the mean richness to obtain more realistic BG counts. Error bars are
assumed to be the Poissonian errors.  The figure shows as the BG
counts are higher in DARC clusters than in relaxed
clusters. Considering the cluster regions within $R_{200}$, the mean
values for both samples separately are $N_{\rm BG,DARC}=23.1\pm3.1$
and $N_{\rm BG,rel}12.0\pm2.7$ for DARC and relaxed
clusters. Therefore, DARC clusters contain a factor two more BGs than
relaxed clusters and the difference is significant at the $2.7\sigma$
c.l..

\begin{figure}
\centering 
\resizebox{\hsize}{!}{\includegraphics{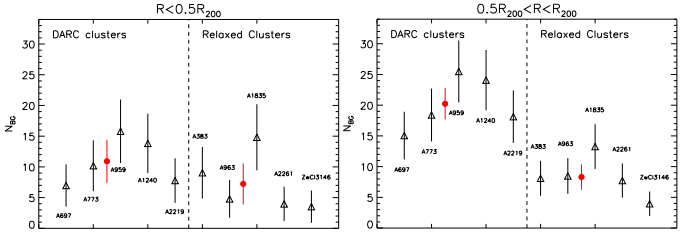}}
\caption{BG counts in the inner region, $R<0.5R_{200}$ ({\em left
    panel}) and in the outer region, $0.5R_{200}<R<R_{200}$ ({\em
    right panel}). Symbols are the same as in Fig.~\ref{fig:BG}.}
\label{fig:BG_radius}
\end{figure}

We also study the spatial distribution of BG galaxies. We compute BG
counts in two separate regions, $R<0.5R_{200}$ and
$0.5R_{200}<R<R_{200}$ (see Fig.~\ref{fig:BG_radius}). For the inner
cluster region, $R<0.5R_{200}$, we obtain $N_{\rm BG,
  DARC}=10.9\pm2.7$ and $N_{\rm BG, rel}=7.2\pm2.4$. For the outer
cluster region, $0.5R_{200}<R<R_{200}$, we obtain $N_{\rm BG, DARC}=
20.2\pm2.5$ and $N_{\rm BG, rel}=8.3\pm2.1$. Thus the difference in BG
counts is restricted to the outer region, where is significant at
$3.6\sigma$ c.l..

We correlate $N_{\rm BG}$ with the redshift and the mass of the
cluster. We do not observe any clear relation in both $N_{\rm BG}-z$
and $N_{\rm BG}-M_{200}$ relations.

Our above results are based on photometric BG counts opportunely
corrected with the average field counts (see Sect.~\ref{sub:lf}). Due to
the poor number statistic of BGs in the individual clusters, we also 
use another approach to check the above results. We also resort to the 
SDSS spectroscopic survey since redshift data can provide the correct cluster 
membership. Unfortunately, SDSS survey is not spectroscopically complete. 
Indeed, within the cluster regions of our sample, SDSS DR8 provide redshifts 
$z$ for 60\% of galaxies having $M_{r'}<-22.5$. Thus, we cannot assess 
the cluster membership for all BGs. Here we consider a mixed approach: 
we counts BGs in the photometric sample and then weight these numbers 
on the base of the spectroscopic information.

As for photometric data, Table~\ref{bg_spec} lists the mean number of
BGs for DARC and relaxed cluster (Cols.~2 and 3); the respective
difference and ratio between $N_{\rm BG}$ in DARC and relaxed clusters
(Cols.~4 and 5).  These values confirm that the strong difference
between DARC and relaxed clusters are due to the outer cluster region,
while nothing can be said about the inner region due to the poor number
statistics. Note that here we do not normalize with the respective
cluster richness as in the above section: the net effect would be of
slightly reducing the observed difference between DARC and relaxed
clusters. Therefore, the results of this section should be 
looked at as being very conservative.

We use the spectroscopic information to weight the above ratios. We 
consider galaxy members those galaxies with a difference of $\delta
z=0.01$ ($\delta z=0.007$) from the cluster redshift, i.e. $\delta cz\sim 
3000$ \kss ($\sim 2000$ \kss) from the mean cluster velocity.
Table~\ref{bg_spec} lists the number of cluster BGs over the number of
all BGs having redshift, $N_{\rm mem}/N_z$, for DARC and relaxed
cluster (Cols.~6 and 7), and the values of the ratios DARC/relaxed 
now weighted with the respective $N_{\rm mem}/N_z$, i.e.
WRatio(DARC/rel)=Ratio(DARC/rel)*[$N_{\rm
    mem}/N_z{\rm(DARC)}]\times[N_{\rm mem}/N_z{\rm (rel)}]^{-1}$
(Col.~8).  When comparing the values of ``WRatio'' with those of
above ``Ratio'' we see that the result of DARC clusters being richer
in BGs than relaxed cluster within $0.5R_{200}$ is confirmed. 

\begin{table*}[!ht]
        \caption[]{BGs in photometric samples weighted with
          spectroscopic samples.}
         \label{bg_spec}
              $$ 
           \begin{array}{l r r c c c c c}
            \hline
            \noalign{\smallskip}

            \hline
            \noalign{\smallskip}  
&\multicolumn{4}{c}{\rm{from\ photometric \ survey}}&\multicolumn{2}{c}{\rm{from \ spectroscopic \ survey}}& {\rm photo+spec}\\   
 & N_{\rm BG, DARC}    & N_{\rm BG, rel}& {\rm Diff.}&{\rm Ratio} & N_{\rm mem}/N_z ({\rm DARC})^{\rm a}   & N_{\rm mem}/N_z ({\rm rel})^{\rm a}&{\rm WRatio}^{\rm a}\\   
\hline
R<R_{200}   & 65 & 45 & 20\pm10 &1.4& 10/41(8/41) & 8/25(6/25) & 1.1(0.9) \\
R<0.5R_{200}& 13 & 16 &  3\pm 6 &0.8& 4/13(4/13)  & 4/8(4/8)   & 0.5(0.5) \\
R>0.5R_{200}& 52 & 29 & 23\pm 9 &1.8& 6/28(4/28)  & 4/17(2/17) & 1.6(2.2) \\ 
                        \noalign{\smallskip}			    
            \hline					    
            \noalign{\smallskip}			    
            \hline
	    					    
         \end{array}
     $$ 
\begin{list}{}{}
\item[$^{\rm{a}}$] Where member galaxies are those  
having $\delta z =0.01$ 
($\delta z =0.007$).
\end{list}
\end{table*}

\section{BGs Luminosity}
\label{l}

We investigate whether the contribution in luminosity of BGs is
different between DARC and relaxed clusters.
We compute the ``global'' luminosity for $M_{\rm r'}<-20$ using the 
  counts of Sect.~\ref{lf}, i.e.:

\begin{equation}
L_{\rm global}=\sum_i^N N_i(m)l_i(m), 
\end{equation} 

\noindent where the sum is performed over all the $N$ magnitude bins
with galaxy number $N_i(m)$ and mean luminosity $l_i(m)$. The
transformation from absolute magnitudes to absolute luminosity in
units of solar luminosities is performed using the solar magnitude
$M_{r\odot} = 4.62$. The luminosity associated to BGs, $L_{\rm BG}$, is
computed in the same way out to $M_{\rm r'}<-22.5$. Table~\ref{tab:tabLM} 
lists the luminosities for each cluster. Values within parenthesis are 
errors estimated at $1\sigma$ c.l.. 


\begin{table}[!ht]
        \caption[]{Luminosity content of BGs.}
         \label{tab:tabLM}
              $$ 
           \begin{array}{l c c}
            \hline
            \noalign{\smallskip}
            \hline
            \noalign{\smallskip}

\mathrm{Name} & L_{\rm global} &  L_{\rm BG} \\
              & (10^{12} L_\odot)     &  (10^{12} L_\odot)  \\
\hline
\noalign{\smallskip}  

A697  &  4.38(0.24) &  0.68(0.04) \\
A773  & 10.71(2.33) &  3.43(0.75) \\
A959  &  7.20(2.11) &  2.53(0.74) \\
A1240 & 16.14(4.74) & 10.78(3.17) \\
A2219 & 12.08(2.96) &  3.32(0.81) \\
\hline
\noalign{\smallskip}  
A383     &  3.03(0.63) & 2.90(0.60) \\
A963     &  6.29(0.67) & 0.78(0.08) \\
A1835    & 16.95(2.74) & 7.04(1.14) \\
A2261    & 12.31(2.52) & 2.55(0.53) \\
ZwCl3146 &  6.95(3.52) & 1.74(0.09) \\  	  
            \noalign{\smallskip}			    
            \hline					    
            \noalign{\smallskip}			    
            \hline					    
         \end{array}
     $$ 
\end{table}


Note that $L_{\rm BG}$ represents a non-negligible fraction of the
global luminosity for $M_{\rm r'}<-20$ and the ten clusters span over
a wide range of luminosities. We observe two dramatic cases, these are
A1240 in the DARC cluster sample and A383 in the relaxed sample, where
the luminosity fraction associated to the BGs, $f_{\rm L,BG}$, are about 
67\% and 96\% of the total luminosity, respectively. Considering the whole 
samples, we find that the mean luminosity fractions of BGs are $f_{\rm
  L,BG,DARC}=0.35\pm0.19$ and $f_{\rm L,BG,rel}=0.39\pm0.33$ for
DARC and relaxed clusters. If we do not consider the two extreme 
cases of A1240 and A383, we find $f_{\rm L,BG,DARC}=0.27\pm0.09$ and
$f_{\rm L,BG,rel}=25\pm0.13$. From these values, we conclude that
DARC and relaxed clusters do not significantly differ for the
luminosity content associated to their BGs.

\section{Discussion \& conclusions}
\label{disc}

We compare five unrelaxed clusters (DARC sample) with relaxed
clusters. As far as possible we have potential biases under control. 
The relaxed clusters are similar to unrelaxed clusters for their
redshift range, X-ray temperature, and mass (see Tables~\ref{tabdarc}
and ~\ref{tabrelax}). Moreover, we always work within physical rather
than fixed radii.

We find that very unrelaxed are comparable to relaxed clusters
for the statistical properties, i.e. the parameters of the LF and the
relative behaviour between red and blue galaxies. In particular,
we agree with  a series of previous results.

We compare $M^*$ and $\alpha$ parameters of the composite Schechter
luminosity functions with results obtained by Popesso et
al. (\cite{pop06}). Transforming their $M^*$ estimates (see their Tables 
1 and 2) to our cosmology, applying the factor $5\log(h_{70})$, 
we obtain values in perfect agreement.


%
%

In addition, in agreement with Popesso et al. (\cite{pop06}, their Fig.~10) 
and de Filippis et al. (\cite{def11}, their Fig.~11) the LF profile 
does not depend on the considered cluster region within $M_r<-19$.

We also find evidence of the variation of the blue fraction of
galaxies with redshift.  It is very noticeably that we can detect the
effect even if our study is limited to a small redshift range. In
particular, note that extrapolating our $f_B(z)$ relation to z=0.35, we
obtain a $f_B(z=0.35)=0.34\pm0.05$. This estimate well agrees with those
obtained by Andreon et al. (\cite{and06}). They find $f_B(z=0.35)=0.33\pm0.05$
with clusters located in a redshift range $0.3-0.4$. When
extrapolating to $z=0.5$, we obtain $f_B(z=0.5)\sim0.5$. This finding is in
very good agreement with cosmological galaxy formation models for
cluster with similar mass (see Menci et al. \cite{men08}, their
Fig.~3). 

Our new first result is that relaxed clusters contain fewer BGs and
these BGs are more concentrated in cluster inner regions.  On the
contrary, unrelaxed clusters present more BGs and more homogeneously
distributed within the whole cluster. The BG richness of DARC clusters
is the robust and significant result of Sect.\ref{bg}.

Our result is not in contrast with that of de Propris et
al. (\cite{dep03}) who found similar LF between substructured and non
substructured clusters since: i) their study is devoted to LF
parameters, for which we find no difference, too; ii) our sample of
DARC clusters are likely to be much more far from dynamical
equilibrium than their sample of substructured clusters. In fact, DARC
clusters are found to be cases of major, very important mergers, in
agreement with the rarity of the diffuse radio sources phenomena
($\sim 10\%$ of clusters, Giovannini \& Feretti \cite{gio02}).
Moreover, DARC clusters should all be recent mergers since radio
emissions are expected to have a short life time, i.e. of the order of
a few $10^8$ years (e.g., Giovannini \& Feretti \cite{gio02}; Skillman
et al. \cite{ski11}), in agreement with times estimated in a few
individual clusters (e.g., Markevitch et al. \cite{mar02}; Girardi et
al. \cite{gir08}).

Our second, new result is that the luminosity content in BGs is the
same in DARC and relaxed clusters. The consequent scenario is that
the (more numerous) BGs lying in the outer regions of merging clusters
will merge to form (more luminous) BGs in the inner regions of relaxed
clusters.  This combined formation/evolution of BGs and the parent
clusters well agree with the results of Lin \& Mohr (\cite{lin04}) who
analyzed the correlation between BCG luminosity and parent cluster
mass, supporting a recent formation of the brightest galaxies in
the context of the hierarchical scenario.

We plan to extend our study to a larger sample of DARC clusters,
by examining their dynamics to compute more reliable mass
estimates. Another possible extension of this work is looking at
deeper magnitudes thus to investigate possible effects on the faint 
end of the galaxy LF connected with the cluster evolution.

\begin{acknowledgements}

We are in debt with Andrea Biviano for his useful comments on this
work. We also thanks for the creation and distribution of the SDSS
Archive, provided by the Alfred P. Sloan Foundation and other
Participating Institutions. The SDSS Web site is
\verb+ http://www.sdss.org/+.  This work has been funded by the
Spanish Ministry of Science and Innovation (MICINN) under the
collaboration grants AYA2010-21887-C04-04 and AYA2010-21322-C03-02. MG
acknowledges financial support from PRIN-INAF-2010.

\end{acknowledgements}

\end{document}